\documentclass[reprint, superscriptaddress]{revtex4-2}

\usepackage{graphicx}
\usepackage{url}
\usepackage{amsmath,enumitem}
\usepackage{mathtools}
\usepackage{amssymb}
\usepackage[dvipsnames]{xcolor}
\usepackage{seqsplit}
\usepackage{url}

\usepackage{lipsum}  
\usepackage{hyperref}
\usepackage{nameref}
\usepackage{booktabs}
\usepackage{float}
\usepackage[normalem]{ulem}

\usepackage{xr}

\makeatletter
\newcommand*{\addFileDependency}[1]{
  \typeout{(#1)}
  \@addtofilelist{#1}
  \IfFileExists{#1}{}{\typeout{No file #1.}}
}
\makeatother

\begin{document}

\title{Vector field theory in motion: Revealing latent potentials in soccer dynamics\\}

\author{Paula Rodr\'{\i}guez-S\'anchez}
\affiliation{Complex Systems Group and G.I.S.C, Universidad Rey Juan Carlos, M\'ostoles, 28933, Madrid, Spain}

\author{Roberto López del Campo}
\author{Ricardo Resta}

\affiliation{Football Intelligence \& Performance de LALIGA, LaLiga, Torrelaguna 60, 28043 Madrid, Spain}

\author{José J. Ramasco$^*$}
\affiliation{Instituto de F\'isica Interdisciplinar y Sistemas Complejos IFISC (CSIC-UIB), Campus UIB, 07122 Palma de Mallorca, Spain}

\author{Javier M. Buldú}
\affiliation{Complex Systems Group and G.I.S.C, Universidad Rey Juan Carlos, M\'ostoles, 28933, Madrid, Spain}

\author{Johann H. Martínez}
\affiliation{Complex Systems Group and G.I.S.C, Universidad Rey Juan Carlos, M\'ostoles, 28933, Madrid, Spain}

\email[]{Email: jramasco@ifisc.uib-csic.es}

\begin{abstract}
\section*{Abstract}
Team sports offer a natural laboratory for studying collective motion in competitive environments. While recent tracking technologies enabled detailed analysis of player and ball trajectories, most existing approaches rely on discrete events and local kinematic descriptors. This limits the possibility of studying team collective behavior. Here we introduce a physics-based framework to characterize collective dynamics using empirically reconstructed velocity vector fields defined over the pitch. Taking soccer as a paradigmatic example, we show that both ball and player movements generate smooth, predominantly irrotational vector fields. Each of these fields is generated by a potential providing compact, physically interpretable representations of collective motion and encoding how teams control, or constrain play across space during attacking and defensive phases in each match. Applied to an entire season of top-level competition, the resulting potentials reveal stable team-specific fingerprints and quantitative associations between potential gradients and effective ball progression.

\end{abstract}


\maketitle 
\section*{Introduction}

Over the last two decades, physics has increasingly contributed to the quantitative understanding of sports as complex dynamical systems \cite{goff2013,balague2013,spearman2017,buldu2018,galeano2022}. Advances in data acquisition technologies, together with concepts originating in statistical physics, nonlinear dynamics, and complex systems theory, have enabled the study of collective behavior in competitive environments. Team sports, and soccer in particular, have emerged as natural laboratories for investigating collective motion, coordination, and decision-making under constraints \cite{Gudmundsson2017,Memmert2017,marcelino2020,burriel2021}.

A major contribution of statistical physics to sports analytics has been the shift from isolated event-based descriptions toward continuous, spatiotemporal representations of play. Early approaches focused on descriptive statistics and notational analysis, but the availability of high-frequency tracking data has enabled the modeling of teams as interacting systems of agents whose dynamics unfold on the pitch \cite{Gudmundsson2017}. Within this framework, soccer matches can be studied as high-dimensional dynamical processes, where collective patterns emerge from local interactions among players, the ball, and tactical constraints \cite{marcelino2020,raabe2023,novillo2024,novillo2024b,dasilva2025}.

One influential line of research has applied network science to soccer, modeling teams as passing networks in which players are nodes and passes define weighted, directed edges \cite{Duch2010,buldu2018}. These studies revealed robust structural properties of team organization, such as small-world characteristics, heterogeneity in player centrality, and the relationship between network topology and team performance \cite{buldu2019,alves2025}. Extensions incorporating temporal and multilayer representations further demonstrated how network structure evolves during different phases of play and competitive contexts \cite{novillo2024}. Complementary research has focused on the kinematics of players on the pitch. Studies of collective motion have identified regularities in team centroids, surface area, stretch indices, and synchronization patterns, revealing how teams coordinate spatially during attacking and defensive phases \cite{frencken2008,vilar2012}. Other works have analyzed individual and collective displacement statistics, showing that player trajectories exhibit strong contextual dependence \cite{bialkowski2014,gonccalves2018}. 

Despite these advances, most trajectory-based analyses treat motion as a sequence of displacements rather than as a continuous field defined over space. In contrast, field-theoretical approaches, widely used in physics to describe fluids, electromagnetic phenomena, and collective transport, offer a natural framework for capturing how local velocity patterns give rise to global structure. Recently, such approaches have been successfully extended beyond physical systems to model human mobility in urban environments. In particular, Mazzoli al. introduced a vector-field description of recurrent human movement, showing that aggregated displacements give rise to robust human mobility fluxes and associated scalar potentials that encode latent mobility preferences \cite{mazzoli2019,liu2024}. These works demonstrated that human mobility can be characterized at a mesoscale level, between individual mobility and the description of full urban areas, with fields and potentials encoding the basic symmetries at play. The field description is valid to differentiate between the outcome of mobility models since it offers a view that captures at the same time the direction and flows of movements, which allowed to close a decade-long controversy on the best performing model \cite{mazzoli2019}. This framework has been also extended to the grouping of individual trajectories \cite{Bongiorno2021,liu2024}. Additionally, it has been shown that knowledge of the mobility potential greatly increases the performance of machine learning tools \cite{Shida2020, Shida2022}. Similar ideas have been applied, for example, to pedestrian mobility \cite{Bongiorno2021} and logistics \cite{yang2022,liu2024,Farokhnejad2026}.
A related application of vector calculus to soccer was introduced by Morishita et al. \cite{Morishita2025}, who constructed last-pass vector fields and derived scalar and vector potentials to reveal typical passing flows and tactically meaningful deviations from them. Importantly, these vector fields were obtained from event data containing only the coordinates of the starting and ending points of each pass, rather than from the ball's actual movement trajectory.

In this paper, we build on this methodological framework and adapt it to the context of professional soccer. Unlike urban mobility, soccer dynamics occur in a bounded, highly structured, and competitive environment, where interactions are instantaneous and strongly constrained by tactical objectives. By reconstructing empirical velocity vector fields from high-resolution tracking data, we treat players and the ball as agents moving within an embedded spatial domain. This approach focuses on demonstrating whether the motion in soccer can be described by conservative vector fields and whether meaningful scalar potentials can be derived. With this aim, we analyze tracking data from $100$ matches of top-level professional soccer to reconstruct velocity fields associated with ball movement and player displacement during attacking and defensive phases. We first introduce the empirical construction of these fields and examine their spatial organization. We then test the conditions required for the existence of scalar potentials by evaluating the curl and divergence of the reconstructed fields. Upon confirming their conservative nature, we compute the corresponding scalar potentials and show that they provide a compact representation of collective dynamics. Finally, we demonstrate that these potentials act as team-specific fingerprints, enabling quantitative comparisons that may reveal latent tactical structures that are not directly observable from raw trajectories or traditional performance metrics.

\section*{Results}
\subsection*{Empirical velocity vector field \texorpdfstring{$\vec{\mathbf{V}}$}{V}}
We used high-resolution tracking data from 100 soccer matches of $20$ professional soccer teams playing the first ten rounds of the first division of the Spanish National League ({\it Laliga}) during the season 2019/2020. 
We reconstructed three different velocity fields for each team throughout the season: one for the ball in possession, one for the players during the attacking phases, and one for the players during the defensive scenario.
Because the dimensions of the playing fields may vary, we standardized all of them by mapping their domains onto a square space of $L=100$ dimensionless units $(\eta)$.
In all cases, the pitch is discretized into a grid of $m \times m = 20 \times 20$ regions resulting in $n=400$ equally sized cells of length $5~\eta$, each. This setup allows us to track when an agent, e.g., the ball or a player, crosses the $i$-th cell at a given time $t$ (see the Section Methods for details).
For each crossing $j$ of the $i$-th cell, we recorded the corresponding instantaneous velocity vector $\vec{\boldsymbol{u}}_{ij} = u_{ij} \hat{\boldsymbol{u}}_{ij}$, where $u_{ij}$ is the speed and $\hat{\boldsymbol{u}}_{ij}$ is the unit vector indicating the direction of motion.
The accumulated velocity vector associated with cell \( i \) is defined as:
\begin{equation}
\vec{\boldsymbol{u}}_i = \sum_{j=1}^{N_i} \vec{\boldsymbol{u}}_{ij},
\end{equation}
where \( N_i \) is the total number of crossings in cell \( i \) during a match.
To obtain the normalized velocity field, we divided by \( N_i \), yielding:
\begin{equation}
\boldsymbol{V}_i = \frac{\vec{\boldsymbol{u}}_i}{N_i} = \sum_{j=1}^{N_i} \frac{u_{ij}}{N_i} \hat{\boldsymbol{u}}_{ij}.
\end{equation}
This normalization corrects for biases arising from differences in the number of events per cell, enabling a meaningful comparison of velocity patterns across the pitch regardless of local event density.
\begin{figure}
\begin{centering}
\includegraphics[width=0.46\textwidth]{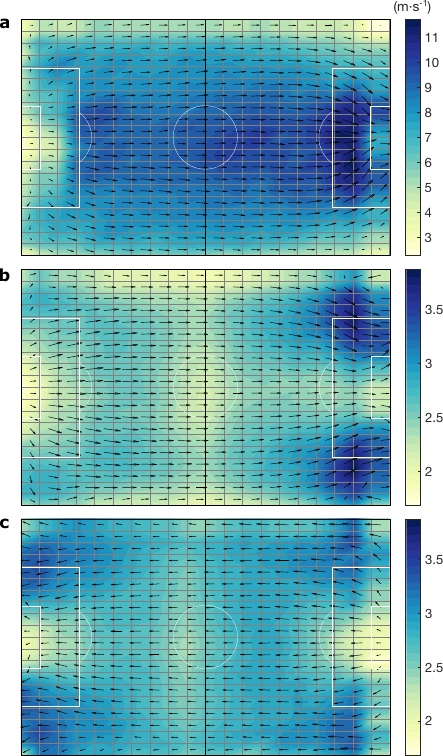}
\par\end{centering}
    \caption{\textbf{Velocity fields.} Each subfigure shows a soccer pitch divided into 400 square cells. Velocity vectors from 100 matches are aggregated within each cell. The colormap represents the magnitude of the average speed in ($\mathrm{m{\cdot}s^{-1}}$). \textbf{(a)} The vector field derived from ball velocity is higher in magnitude than that of \textbf{(b)} players' velocity field during the attacking phase, and  \textbf{(c)} players' movement during the defensive phase.}
    \label{fig:vector_field}
\end{figure}

Figure \ref{fig:vector_field}\textbf{a} shows the velocity field of the ball with $N_i$ as the total number of crossings in the $i$-th cell for all teams and matches.
To standardize the data, we assumed a left-to-right direction of play for the team under analysis, with the opponent's goal on the right. For teams originally attacking in the opposite direction, we inverted the field orientation. Specifically, we reversed the position and velocity vectors of all players and the ball so that the team is now attacking toward the right goal and defending the left. Under this approach, we obtained a unique velocity field for the ball, showing how vectors originated from a common point---the goal area in the defensive third---and extended toward the opponent's area. 

We associated the color intensity of each cell with the magnitude of the average speed, $\overline{u_i}$. In this way, we can observe how the ball's speed increases when the team initiates a play in its defensive third, and continues to escalate as it approaches the opponent's penalty area. Evidently, ball speed decreases along the edges of the field, including the goal areas. The maximum average ball speed is approximately $11\,\mathrm{m{\cdot}s^{-1}}$. While this value appears lower than the reported ball velocities just after instep and side-foot kicks—typically around $23\, \mathrm{m{\cdot}s^{-1}}$ and $28\, \mathrm{m{\cdot}s^{-1}}$, respectively \cite{maly2014,nunome2002}—it remains consistent, as we consider average speeds across all matches, which naturally smooths peak ball velocities.

Maximum average player speeds are approximately $4\,\mathrm{m{\cdot}s^{-1}}$ during both attacking (Fig.~\ref{fig:vector_field}\textbf{b}) and defensive (Fig.~\ref{fig:vector_field}\textbf{c}) phases, which are in line with reported player's velocity profiles~\cite{mizelman2024}.
However, key differences emerge in the spatial structure of players' velocity fields compared to the ball's ones. 
Ball velocity field amplifies its magnitude facing the goal area, while players' fields intensify when approaching the opponent's goal from the corners of the offensive third. 
Regarding the defensive phase, all four corners exhibit intensified player speeds. Notably, the right-side corners are associated with higher speed values, as players accelerate by changing their velocity directions. Meanwhile, the left-side corners cover broader areas of speed changes as consequence of either players' efforts to contain the opponent's positional advantage or simply their deceleration.
In view of all, the spatial distribution of ball velocity profiles as well as player dynamics suggests the existence of a mesoscopic structure capable of generating the observed velocity fields. As a consequence, Field Theory appears to be an alternative mathematical framework to  characterize some aspects of a team's tactical fingerprint.
\begin{figure*}[htb]
\begin{centering}
\includegraphics[width=1\textwidth]{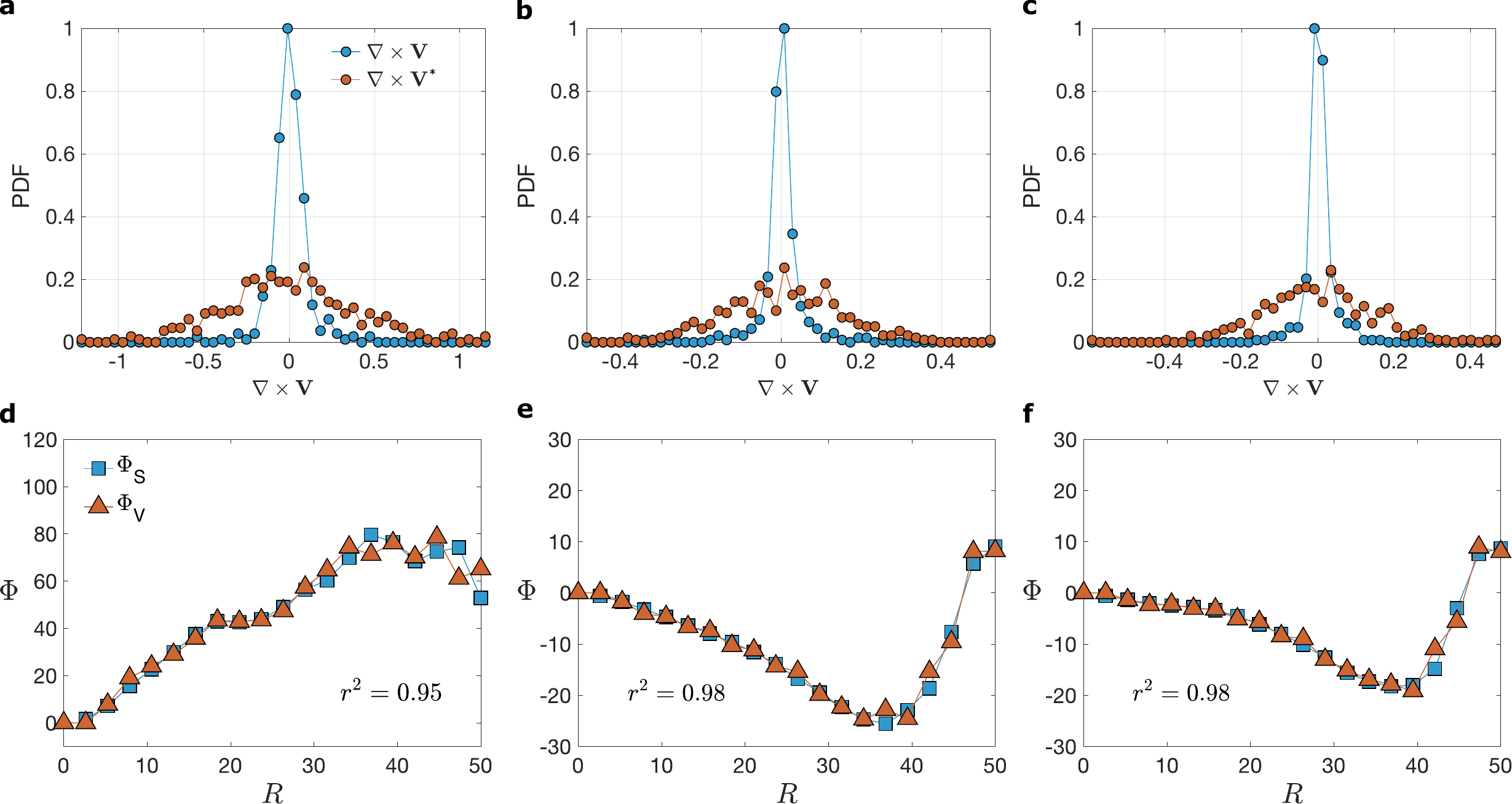}
\par\end{centering}
    \caption{\textbf{Curls and Gauss's theorem.} Upper panel: Discrete PDF of curls for empirical $\boldsymbol{V}$ and null models $\boldsymbol{V^*}$ corresponding to (\textbf{a}) ball movement, (\textbf{b}) player movement during the attacking phase, and (\textbf{c}) the defensive phase. Blue dots indicate the PDF of the empirical fields, and orange dots correspond to that of the null model.
    Bottom panel: Evaluation of Gauss's theorem for the same cases at different spatial scales $R$. Blue squares indicate the total surface flux $\Phi_S$, and orange triangles denote the volume integral of the divergence $\Phi_V$ for (\textbf{d}) ball movement, and players in the (\textbf{e}) attacking and (\textbf{f}) defensive phases. $r^2$ in each panel denotes the squared Pearson correlation coefficient}.
    
    \label{fig:gauss_curl}
\end{figure*}

\subsection*{On the search of a scalar potential}
The existence of a vector field does not necessarily imply the existence of a scalar potential, except under specific conditions. These are $(1^{st})$ that the field is irrotational and $(2^{nd})$ defined over a simply connected domain. 
Stokes’ theorem establishes a fundamental relationship between the circulation of a vector field along a closed curve $C$, and the curl of that field over the surface it encloses. Specifically, it states that the line integral of a vector field \(\boldsymbol{V}\) along \(C\) is equal to the surface integral of its curl over any surface \(S\) bounded by \(C\):
\begin{equation}
    \oint_C \boldsymbol{V} \cdot d\boldsymbol{r} = \int_S (\nabla \times \boldsymbol{V}) \cdot d\mathbf{S} .
\end{equation}
Here, \(d\mathbf{r}\) is the differential line element tangent to the curve \(C\), and \(d\mathbf{S} = \hat{\mathbf{n}}\, dS\) is the differential surface element, with \(\hat{\mathbf{n}}\) as the unit normal vector to the surface \(S\). The orientation of \(\hat{\mathbf{n}}\) is determined by the right-hand rule, ensuring consistency with the positive (counterclockwise) orientation of the boundary curve \(C\). In physical terms, the line integral represents the circulation of the vector field, which may be nonzero when the curl is nonzero. Conversely, if \(\nabla \times \boldsymbol{V} = \mathbf{0}\) throughout a simply connected domain, the field is said to be irrotational. Under this condition, \(\boldsymbol{V}\) can be expressed as the positive gradient of a scalar potential \(\phi\), i.e., \(\boldsymbol{V} = -\nabla \phi\), implying the existence of a conservative potential field.

To ensure the existence of $\phi$, its domain must be simply connected, free of holes and discontinuities. 
We verify this via the divergence theorem, also know as Gauss’ theorem in a physical context, which is expressed as:
\begin{equation}
    \oint_S \boldsymbol{V} \cdot d\boldsymbol{S} = \int_{\text{V}} (\nabla \cdot \boldsymbol{V})\, dv,
    \label{eq:gauss}
\end{equation}
where \(\nabla \cdot \boldsymbol{V}\) denotes the divergence of the vector field \(\boldsymbol{V}\), and \(dv\) is the differential volume element. 
The left-hand side of Eq.~\ref{eq:gauss}., reads for the total flux \(\Phi_S\) of \(\boldsymbol{V}\) through a closed surface \(S\) which is equal to the net flux \(\Phi_V\) generated within the volume it encloses, as quantified by the divergence of \(\boldsymbol{V}\). In the context of a two-dimensional vector field, Eq.~\ref{eq:flux_surface} represents the net flux,
\begin{align}
    \Phi_S &= \oint_l \boldsymbol{V} \cdot \hat{\boldsymbol{n}}\, dl, \label{eq:flux_surface} \\
    \Phi_V &= \int_S (\nabla \cdot \boldsymbol{V})\, dS. \label{eq:flux_volume}
\end{align}
across a closed curve $l$, where $\hat{\mathbf{n}}$ is the outward-pointing normal vector and $dl$ is the differential length along the curve. $\Phi_S$ is then compared to $\Phi_V$, the surface integral (Eq.~\ref{eq:flux_volume}), taken over the area of the surface $S$ enclosed by the curve $l$. In general, the amount of field flowing out through a closed perimeter ($\Phi_S$) equals the total divergence of the field within the enclosed area ($\Phi_V$).
\begin{figure*}[hbt]
\begin{centering}
\includegraphics[width=1\textwidth]{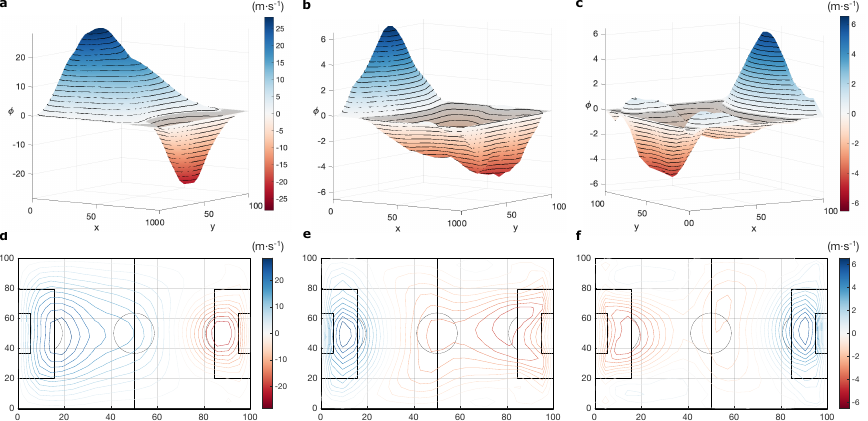}
\par\end{centering}
    \caption{\textbf{Potential $\mathbf{\phi}$ and equipotential lines in the season.} Panel \textbf{a} represents the potential field generated by the ball’s dynamics, while \textbf{d} captures its corresponding equipotential lines. Panel \textbf{b} and \textbf{c} show the potential derived from the movement of attacking and defending players, respectively. \textbf{e} and \textbf{f} show the corresponding contour lines for each of these cases.
    The color bar indicates the potential magnitude, with blue regions representing peaks and red regions indicating sinks.}
    \label{fig:potential}
\end{figure*}
Figure~\ref{fig:gauss_curl} displays the empirical computations of the curl (upper panel) and Gauss’ theorem (lower panel) for vector fields $\boldsymbol{V}$ presented in Fig.~\ref{fig:vector_field}. The curl was calculated via central finite differences applied directly to the two-dimensional vector field data, yielding a scalar field oriented along the $z$-axis. A discrete probability density function (PDF) of the curl across all grid cells was then constructed. To validate these results, the procedure was repeated using a null model $\boldsymbol{V^*}$, where vector directions were randomized within each cell while preserving their magnitudes. Although an analytically irrotational field satisfies \(\nabla \times \boldsymbol{V} = \mathbf{0}\), numerical approximations on discretized, bounded domains introduce errors in the finite-difference estimates of the curl. Moreover, empirical data are subject to noise, which may be amplified when calculating derivatives. These factors notwithstanding, they do not preclude the detection of a conservative field.

Figure~\ref{fig:gauss_curl}\textbf{a} shows the PDF of cell-wise curl values for the empirical vector field $\boldsymbol{V}$ of the ball (blue) and its corresponding null model $\boldsymbol{V^*}$ (orange). The distribution of $\nabla \times \boldsymbol{V^*}$ is mostly near the origin, while that of $\nabla \times \boldsymbol{V}$ sharply peaks at the origin, displaying a higher probability of cell-wise curl around zero. Figures~\ref{fig:gauss_curl}\textbf{b} and~\ref{fig:gauss_curl}\textbf{c} present analogous results for the vector fields of players during attacking and defending phases, respectively. In all cases, the numerical values of the curl are close to zero, indicating a velocity field predominantly irrotational at the aggregated spatio-temporal scale considered here, and therefore with the feasibility of computing a scalar potential for each configuration.

We evaluated Gauss’s theorem using two different closed curves—a circle of radius $R$ and a square of side length $L = 2R$—to test the robustness of the results. For each case, we computed both $\Phi_S$ and $\Phi_V$ across multiple spatial $R$ scales, ranging from 0 to 50 $\eta$, the latter corresponding to the full $L=2R=100~\eta$ length of the field.
The bottom panel of Fig.~\ref{fig:gauss_curl} displays the evaluation of Gauss' theorem for circle curves at different scales. (see Supplementary Figure 1 for the evaluation using square curves.) 
In this vein, Fig.~\ref{fig:gauss_curl}\textbf{d} shows the agreement of both fluxes for the ball movement, Fig.~\ref{fig:gauss_curl}\textbf{e-f} for players in attacking and defensive phases, respectively.
To quantify this agreement, we computed the Pearson correlation coefficient between $\Phi_S$ and $\Phi_V$ across the scales. High correlations indicate that $\boldsymbol{V}$ preserves its divergence structure consistently, supporting the robustness and conservative nature of the reconstructed fields.
Rather than focusing on the absolute values of fluxes, the key insight lies in the agreement of both integrals independently of the scale being evaluated, which confirms the continuity of the spatial domain. 
In this way, we capture not only the local consistency of the vector field but also the regularity of the domain. A strong agreement between $\Phi_S$ and $\Phi_V$ across increasing values of $R$ confirms the absence of discontinuities or holes that would violate the assumptions of the divergence theorem.

\subsection*{The potential of a team}
We numerically computed \(\phi\) as the effective scalar potential for all $i$-th cells with field contribution \(\boldsymbol{V}_i\) by using the definition \(\boldsymbol{V} = -\nabla \phi\) via central finite differences. See Methods for details.
Then we plotted \(\phi\) as a surface over the normalized area in   
Figure~\ref{fig:potential}\textbf{a} displaying the scalar potential associated with ball movement. As spatial derivatives are computed over a dimensionless standardized grid, the scalar potential \(\phi\) must retain the same physical units as velocity.
A pronounced global maximum is observed at the front of the team's own penalty area, indicating a source region from which the vector field diverges. This suggests that ball trajectories tend to originate from this zone (see Fig.~\ref{fig:vector_field}\textbf{a}), consistent with its functional role in soccer as the area where a play is often started.
Conversely, a global minimum is identified near the opponent's goal, particularly in regions where the vector field magnitude increases. This aligns with the expected behavior in attacking phases, where ball speed typically rises as the ball approaches the opposite area. It is important to highlight that zero values of the potential do not imply a vanishing vector field, but rather indicate locations where the flux changes direction or intensity across space.
As a complement, Figure~\ref{fig:potential}\textbf{d} shows the equipotential lines associated with the scalar potential field. These lines are colored from blue (near the global maximum) to red (near the global minimum), providing a visual map of the potential landscape that underlying the velocity field.

The scalar potentials derived from player movements during attacking (Fig.~\ref{fig:potential}\textbf{b}) and defensive phases (Fig.~\ref{fig:potential}\textbf{c}) also exhibit well-defined structures. Although a global maximum and minimum are reported in all cases, the potential amplitudes are lower than those observed for the ball, reflecting the lower velocities of players compared to the ball. Interestingly, while the ball potential presents two pronounced maxima and minima, the potentials associated with player dynamics show more subtle variations, particularly in the location of the minima. Specifically, the scalar potential exhibits two local minima during the attacking phase (Fig.~\ref{fig:potential}\textbf{b}). As shown in Fig.~\ref{fig:potential}\textbf{e}, the equipotential lines become closer near the attacking goal, indicating stronger changes of speeds in this region. The minimum is slightly shifted to the left, which may be associated with higher approach velocities of left-sided attacking players as they advance toward the goal.

Defensive potential (Fig.~\ref{fig:potential}\textbf{c}) exhibits two local minima when players retreat toward their own goal corresponding to areas with high velocity field contributions, one close to the center of the pitch and the other inside their own penalty area (Fig.~\ref{fig:vector_field}\textbf{c}). In addition, two local maxima emerge toward the lateral areas of the defending half. This spatial configuration may reflect the behavior of defensive players, i.e., keeping structured lines when returning, with higher velocities directed toward the central areas of the pitch. As they approach their own goal, defenders tend to preserve positions within central channels rather than drifting toward the flanks.

\subsection*{Global extrema and the gradient of teams' potential}

Obtaining the scalar potential simplifies the analysis of relevant vector field phenomena, as algebraic manipulations are generally more direct with scalars than with vectors.
To further analyze the properties of the potential across teams, we identified the \textit{global maximum} and \textit{global minimum} of the scalar field as
\[
\phi_{\mathrm{max}} = \phi(\vec{r}_{\mathrm{max}}), \quad \phi_{\mathrm{min}} = \phi(\vec{r}_{\mathrm{min}}),
\]
where $\vec{r}_{\mathrm{max}}$ and $\vec{r}_{\mathrm{min}}$ denote the dimensionless grid units $\eta$ at which these extrema occur (see Supplementary Figures 11--13 for the spatial distribution of global maxima and minima).
We then define the \textit{mean gradient} along the straight line connecting these points as a measure of the overall slope of the potential,
\[
m = \frac{\Delta\phi}{\delta},
\]
with $\Delta\phi = \phi_{\mathrm{max}} - \phi_{\mathrm{min}}$, and $\delta = \left\|\vec{r}_{\mathrm{max}} - \vec{r}_{\mathrm{min}}\right\|$. This quantity, $m$, has units of meters per second per grid unit and provides a global estimate of the potential's variation relative to spatial displacement. It also serves as a useful metric to quantify the intensity of the vector field along the direction connecting the potential's extrema.

\begin{figure}
\begin{centering}
\includegraphics[width=0.42\textwidth]{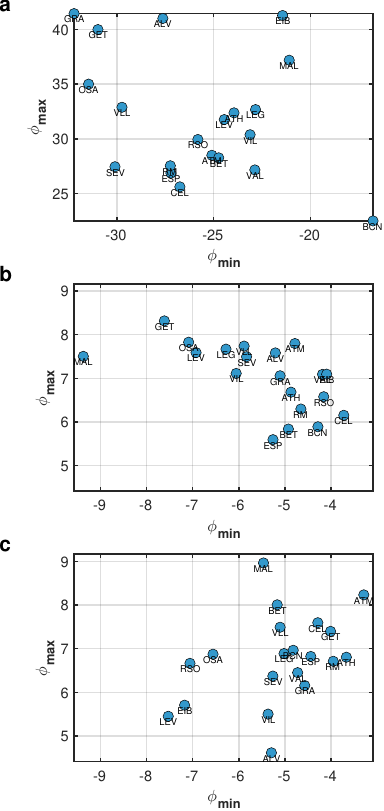}
\par\end{centering}
    \caption{\textbf{Relationship between global maxima and minima of the potential $\phi$ across teams.} 
    Each pair $( \phi_{min},\phi_{max} )$ corresponds to one team potential accounting their matches. \textbf{(a)}, Potential derived from the ball movement. \textbf{(b)} and \textbf{(c)} are the extreme values of the potential generated by players during attacking and defensive phases, respectively.}
    \label{fig:scatter_potential}
\end{figure}

We computed the scalar potentials for all teams to investigate patterns in their extreme values (see Supplementary Figures 2--10 for team-specific potential field computation and comparison between Getafe and Barcelona). As shown in Fig.\ref{fig:scatter_potential}, the relationship between the minimum and maximum potential values is generally nonlinear for the ball and for both phases of the game.

As illustrated in Fig.~\ref{fig:scatter_potential}\textbf{a}, most teams exhibit higher potential values at their own goal ($\phi_{max}$) compared to the opponent's goal ($\phi_{min}$). Although ball speed is typically higher when approaching the opponent's area (Fig.~\ref{fig:vector_field}\textbf{a}), the absolute value of the scalar potential is consistently lower in that region. This suggests that in their own half, teams exhibit more direct and vertically aligned movements toward the front, which results in a more pronounced potential peak. In contrast, offensive sequences near the opponent's goal involve frequent backward and lateral passes. Despite occurring at higher instantaneous speeds, these velocity vectors are less consistently aligned with the primary direction of play, leading to a lower cumulative absolute potential.

Among all teams, Barcelona (BCN) stands out for exhibiting the least extreme potential values, characterized by the lowest maximum and the highest minimum. This configuration suggests slower ball movement in both attacking and defensive zones, indicative of a control playing style that is grounded in ball retention (see Supplementary Table 1 for the complete list of team names and their corresponding acronyms).
In contrast, Granada (GRA) exhibits an opposite behavior having the lowest minimum potential, which is associated with a more direct and offensive playing style. In other words, GRA tends to move the ball faster both at the beginning of the possession and when moving toward the opponent’s area.
An interesting case is Eibar (EIB), which displays a high maximum potential, yet with an intermediate minimum. This indicates that Eibar accelerates the ball the most from its own area, but slows down the velocity as it approaches the opponent's goal, suggesting a potential loss of momentum in the final third of the pitch.

Figure~\ref{fig:scatter_potential}\textbf{b} refers to the players' movement and reveals a similar pattern for BCN. Their players start with lower velocities and arrive at the opposite area also with the lowest speeds compared to other teams, reinforcing the idea of a balanced playing style.
Getafe players, by contrast, also maintain an offensive strategy of achieving high speeds both when restarting play and when approaching the opponent’s area.
However, not all teams keep the same pattern between the ball and the player movement in the offensive phase. Players from Mallorca (MAL) exhibit a high maximum potential in their own area, followed by the lowest minimum potential (highest aligned velocities) when reaching the opponent’s goal. This may indicate that attacking players move faster compared to those of other teams. Interestingly, this pattern is not followed by the ball itself, whose velocities exhibit a distinct behavior, specifically a moderate trend.

Differences in the global minima during the defensive phase are minor across teams, as shown in Fig.~\ref{fig:scatter_potential}\textbf{c}. In contrast, substantial differences are observed in the global maxima, reflecting distinct defensive behaviors in the opponent's half of the pitch. Mallorca exhibits the highest global maximum, reflecting higher velocities when recovering positions toward their own half. Barcelona does not stand out for particularly extreme values, remaining close to the league average during defensive phases. Other teams, such as Levante (LEV) and Eibar, display more pronounced global minima but lower global maximum, suggesting more moderate velocities in the opponent's half and increased defensive pressure as play approaches their own penalty area.
To further assess whether the reported extrema reflect team-specific structure beyond the immediate rival context, we performed a comparison between each team’s potential extrema and a rival-average baseline constructed from the opponents faced in the same matches (Supplementary Fig.~14). In this comparison, the team-specific extrema do not collapse onto the identity line, particularly for ball-related extrema, indicating that the reconstructed potentials are not simply inherited from the rival context.

\begin{figure}
\begin{centering}
\includegraphics[width=0.4\textwidth]{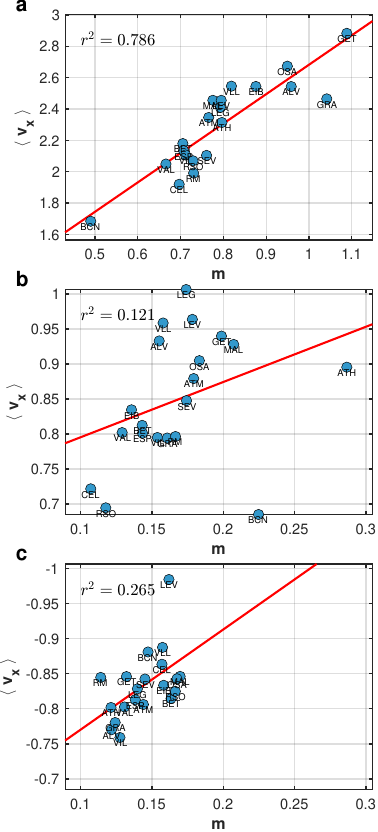}
\par\end{centering}
    \caption{\textbf{Average horizontal speed and field gradient.} Each blue point $(\left\langle v_x \right\rangle, m)$ represents a single team with values aggregated across all matches. Red
    line is the linear fit of data from 
    \textbf{a}, Ball trajectories,
    and players’ motion during
    \textbf{b}, Offensive and, 
    \textbf{c}, Defensive phases.
    }
    \label{fig:scatter_slope}
\end{figure}

Next, we explored how the effective velocity of the ball and the players, moving from one goal to the opposite area, varies in terms of the spatial variation of the potential. To do so, we plotted the average of the horizontal component of velocity $\left\langle v_x \right\rangle$ against the mean gradient $m$ of potentials, for each team. 
Figure~\ref{fig:scatter_slope}\textbf{a} depicts a positive trend between $\left\langle v_x \right\rangle$ and $m$ for the ball potential with $r^2=0.79$. This suggests that the structure of the potential is intimately related to the alignment of velocity vectors implying a better horizontal coordination of the ball.
This trend is consistent with the style of game of specific teams, e.g., Barcelona, has the lowest $\left\langle v_x \right\rangle$ with a low field intensity $m$ which agrees with a game style where the possession of ball needs of slow player's velocities when arriving toward the opposite goal.
Conversely, teams like Osasuna (OSA), Getafe, Eibar or Granada have a more direct style which is reflected into higher $\left\langle v_x \right\rangle$ associated with stronger field variations.

Figure~\ref{fig:scatter_slope}\textbf{b} shows the trend for players during the offensive phase, with $r^2=0.12$. In this case, there is no linear trend and only 12\% ($r^2=0.12$) of the variability of the horizontal velocity is explained through a linear relation with the field gradients.

Next, we investigated the defensive phase, where players are prone to return to their own area (Fig. \ref{fig:scatter_slope}\textbf{c}). In the defensive phases, the correlation is also low ($r^2=0.27$). The remaining variability is probably linked to factors not included in this data-driven model such as sideways movement of players when restoring their position, specific tactical choices like pressure upon rivals, or man-marking, which are not accounted for the scalar potential.
\begin{figure*}
\begin{centering}
\includegraphics[width=1\textwidth]{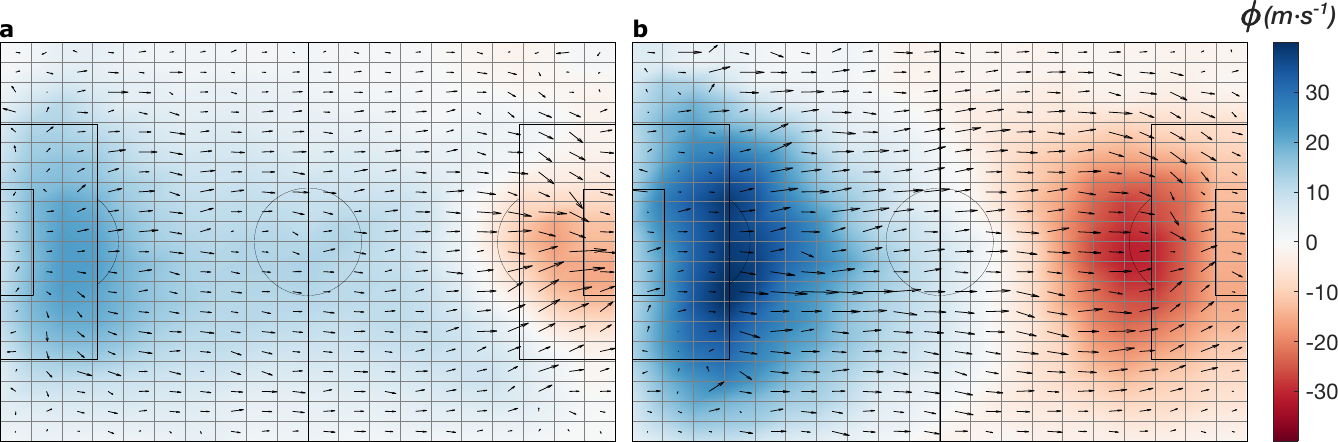}
\par\end{centering}
    \caption{\textbf{$\boldsymbol{V}$ and $\boldsymbol{\phi}$ for Barcelona and Getafe.} 
    Each panel consists of the vector field on each cell with the heatmap representing its associated potential for: 
    \textbf{(a)} Barcelona  and 
    \textbf{(b)} Getafe.}
    \label{fig:team_comparison}
\end{figure*}

\subsection*{The potential as a teams' fingerprint}

The scalar potential is a mathematical function fully retrieved from teams' velocities that encodes their collective dynamics. It provides a powerful framework to characterize and interpret the global structure of the {\it collective velocity}, regardless of whether it refers to the ball or the players. Whereas vector field representations offer a detailed view of velocity vectors, direct comparisons between such fields offer only part of the story. In this regard, the scalar potential acts as an alternative fingerprint of team behavior as it concentrates the complex structure of motion into a continuous scalar representation. Hence, simplifying teams' description and granting global values and mean gradients for uniquely characterize collective playing styles.

Figure \ref{fig:team_comparison}~(\textbf{a}-\textbf{b}) shows the comparison between Barcelona and Getafe, two of the most contrasting teams in the season. By leveraging their ball's vector fields, we derived the corresponding scalar potentials to capture systematic differences across all matches of the season. When examining these vector fields, Barcelona  displays vectors with shorter magnitudes than those of Getafe, while differences among vector's orientations are difficult to discern, restricting vector field comparisons. However, the scalar potentials clearly differentiate the way in which team's dynamics are topographically distributed across the field. 
Although both fingerprints exhibit a spreading area in their own half, and an attractor-like region in the opposite goal, Barcelona's fingerprint is noticeable flatter than that of the Getafe, which is markedly pronounced. 
In other words, Getafe displays highly contrasted regions compared to smoother Barecelona's fingerprint. Consistently, Barcelona's global values are lower than those of the Getafe's counterparts (in $m\cdot s^{-1}$):  $\mathbf{\phi_{max}}=22.499$, $\mathbf{\phi_{min}}=-16.768$, $m = 0.489$, for BCN, and $\mathbf{\phi_{max}}= 39.974$, $\mathbf{\phi_{min}} = -30.981$, $m = 1.089$, for GET. 

The fingerprints also unveil a sort of right tendency for Barcelona to finish plays, something difficult to discern uniquely from vector fields. In contrast, Getafe's fingerprint discloses a bilateral ball movement pattern, specifically when arriving toward opposite goal. 
Barcelona's topographical signature also reveals a broad central area to manage the ball motion, versus the narrow region at the central field when Getafe proposes its play.
Compared to Getafe, Barcelona's potential displays small regions with low velocities when both restarting the play near its own area, and when approaching the rival goal. This is consistent with a team that proposes a controlled and possession-ball style.
Conversely, Getafe's fingerprint shows extended regions of higher velocities near by its own and opposite goals, reflecting a common trait of a direct and offensive playing style.
Altogether, these finding demonstrate that the scalar potential provides a compact and physically interpretable representation of collective motion, enabling quantitative comparisons across teams and revealing alternative playing-style fingerprints that are otherwise hidden in traditional velocity-field analysis.

\section*{Discussion}

In this work, we have introduced a physics framework to characterize collective movement patterns in soccer through empirically derived vector fields defined over the pitch. By reconstructing spatially resolved velocity fields from high-resolution tracking data, we provide a continuous representation of how players and teams traverse space during match play. Our results show that these vector fields capture interpretable signatures of collective behavior, enabling a quantitative description of team dynamics that goes beyond traditional event-based statistics or aggregated kinematic measures.

A central contribution of this study lies in the shift from a discrete, player-centric description of movement to a coarse-grained, field-based representation. This perspective aligns the analysis of team sports with approaches long established in statistical physics and fluid dynamics, where collective motion is naturally described in terms of macroscopic flow fields. The vector fields extracted here encode both the preferred direction and intensity of movement as a function of location, thereby providing a direct link between spatial context and collective intent. Importantly, this representation abstracts away individual identities, allowing team behavior to be analyzed as an emergent property of coordinated interactions rather than as a simple aggregation of individual actions.

Within this framework, the reported team-level values derived from the vector fields acquire a physical and tactical interpretation. The magnitude of the velocity field reflects the intensity with which teams move through different regions of the pitch, while its directional structure reveals systematic tendencies in ball progression, defensive retreat, or lateral circulation. The consistency of these values across matches indicates that they capture intrinsic properties of team behavior rather than transient or opponent-specific effects. At the same time, the observed differences between teams highlight how distinct tactical identities manifest as quantitatively measurable flow patterns. In this sense, the reported values can be understood as macroscopic descriptors of playing style, rooted in collective motion rather than isolated events.

Beyond their descriptive value, these team-specific measurements open the door to principled comparative analyses. Because the vector fields are defined over a common spatial domain, differences between teams can be assessed locally and globally, enabling fine-grained comparisons of how space is exploited or constrained. This also facilitates longitudinal analyses, where changes in the reported values may signal tactical adaptations, contextual effects, or longer-term evolution of team strategies. From a methodological standpoint, the smoothness and spatial continuity of the fields allow for the application of differential operators, such as divergence or curl, which provide additional physically grounded observables related to spatial expansion, compression, or rotational movement.
 
Our approach is also related to, but methodologically distinct from, the vector-calculus framework recently proposed by Morishita et al. \cite{Morishita2025} for the analysis of last-pass events in football. While both studies exploit vector fields and scalar potentials to characterize tactical structure, Morishita et al. focus on a single, discrete event type (the last pass preceding a shot) within the attacking half of the pitch, in a small curated sample of matches, and further decompose the resulting discrete field into gradient and rotational components via a Helmholtz decomposition, associating the rotational residual with defender-induced deviations from the direct route to goal. In contrast, our framework reconstructs continuous velocity fields in the whole pitch from the complete trajectories of the ball and of all outfield players across full matches and an entire season, separately for the attacking and defensive phases of play, and evaluates the conditions for the existence of a single dominant effective scalar potential rather than performing a full Helmholtz decomposition. The two approaches are complementary: combining a Helmholtz decomposition with continuous, tracking-derived fields, or conversely extending the last-pass framework of Morishita et al. to a broader set of tracking-derived events, represent promising directions for future work.

More broadly, the proposed approach complements and extends previous applications of complex systems methods to sports, including network-based representations of passing or positional synchronization analyses \cite{buldu2018,marcelino2020}.
While such methods focus on relational structure or temporal coordination, vector fields explicitly encode motion in space, making them particularly well suited for invasion games, where performance depends critically on coordinated displacement under spatial and temporal constraints. Although the present study focuses on soccer, the methodology is directly transferable to other invasion sports such as basketball, handball, hockey, or rugby, provided that tracking data of sufficient resolution are available. The vector-field formalism also lends itself naturally to further extensions. The reported values could be conditioned on game context, such as scoreline or numerical advantage, or coupled with ball dynamics to study the interaction between player and ball movement. Moreover, vector fields could be integrated with network representations to jointly capture spatial flow and interaction structure, offering a unified view of movement and decision-making. 

At an applied level, these representations provide interpretable yet quantitatively grounded tools that could inform scouting, training design, or the evaluation of tactical interventions. 
Beyond opponent scouting and tactical design, the framework introduced here lends itself to several further applications. First, the magnitude and gradient of the potential could be tracked match by match as compact descriptors of seasonal changes in collective intensity, which may in future work be related to fluctuations in physical condition, tactical rhythm, or match context. Second, the supplementary rival-average comparison suggests a promising longitudinal extension examining whether teams systematically reshape their potential landscapes when facing opponents with profiles similar to those encountered previously, thereby informing opponent-specific preparation. Third, the potential-based fingerprints illustrated in Fig. \ref{fig:team_comparison} provide a compact representation that could support similarity-based classification of teams or playing styles across seasons and competitions.

However, several limitations of the present approach should be acknowledged. On one hand, The construction of smooth vector fields necessarily involves choices regarding spatial discretizations, temporal aggregation, and smoothing, which may attenuate fine-scale or short-lived behaviors. While the reported values are robust at the macroscopic level, rare or highly localized patterns may not be fully captured. Future work could address these issues by adopting multiscale or time-resolved formulations, as well as probabilistic approaches that explicitly quantify variability and uncertainty across matches.
On the other hand, because the vector fields analyzed here are aggregated over full matches and across an entire season, our finding that the curl of these fields is predominantly close to zero (Fig. \ref{fig:gauss_curl}) should be understood as a statement about the season-aggregated, mesoscopic flow, rather than as evidence that football dynamics are irrotational at every instant of play. Local or short-lived rotational structure, potentially relevant during transitions, counter-attacks, or pressing situations, may be attenuated by this aggregation and would require phase- or time-resolved analyses to be properly characterized. Similarly, the agreement we report between the surface flux and the divergence integral (Fig. \ref{fig:gauss_curl}, bottom panels) should be interpreted as a numerical consistency check on the discretized field, rather than as a formal proof of the continuity or simple connectivity of the underlying, continuous velocity field.

Finally, our results contribute to the understanding of collective motion in systems composed of {\it self-propelled agents} (i.e., the players), a central theme in the physics of active matter \cite{bar2020}. Team sports provide a unique empirical setting in which agent-level objectives, interaction rules, and environmental constraints are well defined, yet large-scale behavior emerges in a highly non-trivial manner. By representing player movement through coarse-grained vector fields, we bridge microscopic dynamics and macroscopic flow descriptions, closely paralleling hydrodynamic approaches to active systems \cite{toner2005,marchetti2013}. In this sense, soccer can be viewed as a controlled realization of collective motion far from equilibrium, where coordination, competition, and adaptation continuously shape emergent flow patterns. The framework introduced here therefore not only advances sports analytics, but also offers a data-driven testbed for exploring fundamental questions in active matter physics, including the formation of coherent flows, symmetry breaking in confined domains, and the role of interactions in shaping large-scale transport.

\section*{Methods}
\phantomsection
\label{sec:methods}
\subsection*{Data collection}
The players tracking datasets were obtained from Mediacoach\textsuperscript{\textregistered}, which relies on the Tracab Optical Tracking system \cite{linke2020}, consisting of a multi-camera arrangement operating at a frequency of \(f = \Delta t^{-1} = 25\) frames per second.  Specifically, the system consists of a series of camera sets placed along the stadiums, each set composed, in turn, of three cameras with a resolution of \(1920 \times 1080\) pixels, providing a panoramic stereoscopic view for triangulating both players and the ball. Player positions are reconstructed by estimating the body skeleton and projecting its center onto the pitch. Occasional tracking losses are manually corrected by experienced operators.

The accuracy and validity of the Mediacoach\textsuperscript{\textregistered} datasets have been previously confirmed against GPS-based measurements~\cite{Felipe2019,Pons2019}.  Regarding spatial accuracy, Linke et al.~\cite{linke2020} reported a root mean square error (RMSE) of 9~cm for two-dimensional player positions. Speed estimation accuracy was 0.09~m/s (RMSE), with improved reliability at higher running speeds~\cite{linke2020}.

The dataset analyzed in this work corresponds to matches from the Spanish National League (LaLiga) during the 2019/2020 season. All players who were on the pitch while the ball was in play were considered, ensuring that the analysis reflects periods of active match dynamics. 
Offensive and defensive phases were assigned to each team based on ball possession, as determined by the data provider (Mediacoach\textsuperscript{\textregistered}) through a semi-automated pipeline combining automated event detection with manual supervision by trained analysts. A team was considered to be in its offensive phase in every frame in which it retained possession of the ball, irrespective of the ball's location on the pitch, and in its defensive phase otherwise. No synchronized event data were used in this study. All quantities reported were derived from positional tracking data and the possession labels supplied alongside it.
Following this criterion, 100 matches were analyzed, comprising data from 438 players. Player positions were extracted at the original temporal resolution of 25~frames/s (one frame every 0.04~s). The selected matches correspond to the first 10 fixtures of the season (10 matches per team), yielding a total of approximately 15 million frames for analysis. In addition to tracking data, positional information from the official LaLiga datasets was also incorporated.

\subsection*{Data standardization}
The raw positional data in meters is provided in a physical reference frame \((x,y)\) whose origin is located at the center of the pitch. 
For each match, we accessed the positions of each agent (ball or player) at each equally spaced time frame $t$. 
To obtain reliable estimates of instantaneous velocities we compute a smoothed two-dimensional velocity at each frame. First, by capturing raw frame-to-frame velocities as
$\vec{\boldsymbol{u}}^{\,\text{raw}}_t = \left( \frac{x_{t+1}-x_t}{\Delta t}, \frac{y_{t+1}-y_t}{\Delta t} \right)$, 
with $\Delta t$ the as time interval between consecutive frames.
Then we applied a temporal sliding window of $W=11$ frames to obtain:
\[
\vec{\boldsymbol{u}}^{\,\text{inst}}_t=\;
\frac{1}{W}\sum_{k=1}^{W}
\vec{\boldsymbol{u}}^{\,\text{raw}}_{t+k},
\]
as the smoothed version of the instantaneous velocity in both spatial dimensions.

Pitch dimensions may vary within official regulations. To ensure comparability across matches, we standardized the spatial data by mapping all fields onto a fixed reference scale. Hence, each physical coordinate is mapped onto a square reference frame \((x', y')\) with its origin at the bottom-left corner according to the following parametrization:
\[
x' = \left(x + \frac{l_x}{2}\right)\frac{L}{l_x}, \quad  
y' = \left(y + \frac{l_y}{2}\right)\frac{L}{l_y},
\]
where \(l_x\) and \(l_y\) denote the physical pitch length and width in meters, respectively. Here, \(L=100\) is the characteristic side length of the square field, expressed in dimensionless units \((\eta)\).
This transformation preserves the relative spatial structure while eliminating pitch-to-pitch variability, thus representing all matches on a standardized and isotropic  \(100\times100\) space.

This square was further discretized into a uniform grid of size \(m \times n = 20 \times 20\) blocks, comprising 400 square cells.
The number of velocity-vector observations per cell, $N_i$, ranged from $3300$ to $39138$ (median $18906.5$) for the ball field, from $5579$ to $539420$ (median $179227.0$) for the player fields during attacking phases, and from $5537$ to $625615$ (median $130517.5$) for the player fields during defensive phases. We verified that the main results reported below are qualitatively robust to the choice of grid resolution, repeating the analysis from $2\times 2$ up to $100\times 100$ partitions of the pitch (Supplementary Fig.~15 from Supplementary Information); the $20\times 20$ grid was retained as the resolution offering the best compromise between spatial detail and the number of observations per cell.

Each cell has a side length of 5 \(\eta\), corresponding to an area of 25 \(\eta^2\). In physical terms, each cell covers approximately 0.25\% of the area of a standard pitch. 

This spatial discretization enables us to track the position of agents by detecting their presence within the $i$-th cell at a given time.  
For each crossing $j$ of the $i$-th cell, we record the corresponding instantaneous velocity vector as
\[
\vec{\boldsymbol{u}}_{ij} \equiv \vec{\boldsymbol{u}}^{\,\text{inst}}_{t_{ij}} = u_{ij}\,\hat{\boldsymbol{u}}_{ij},
\]
where $t_{ij}$ denotes the time frame at which the crossing of an agent occurs, 
$u_{ij} = \|\vec{\boldsymbol{u}}^{\,\text{inst}}_{t_{ij}}\|$ is the speed, 
and $\hat{\boldsymbol{u}}_{ij} = \vec{\boldsymbol{u}}^{\,\text{inst}}_{t_{ij}} / u_{ij}$ is the unit vector indicating the direction of motion.

The accumulated velocity vector in cell \(i\) is defined as  
\[
\vec{\boldsymbol{u}}_i = \sum_{j=1}^{N_i} \vec{\boldsymbol{u}}_{ij},
\]  
where \(N_i\) represents the total number of crossings in cell \(i\) during the match.
To obtain the normalized velocity field, the accumulated velocity is divided by \(N_i\), yielding  
\[
\boldsymbol{V}_i = \frac{\vec{\boldsymbol{u}}_i}{N_i} = \sum_{j=1}^{N_i} \frac{u_{ij}}{N_i} \hat{\boldsymbol{u}}_{ij}.
\]
This normalization compensates for differences in the number of crossings across cells, enabling an unbiased comparison of velocity patterns throughout the pitch regardless of local event density.

\subsection*{Vector field computations}\label{sec:curl}

\subsubsection*{Curl computation on the discretized grid.}
Vector field computations were performed in Matlab.
A curl is estimated via partial derivatives 
through finite difference techniques~\cite{zhou1993numerical}. It applies central difference formulas for interior points, while boundary points are handled using forward one-sided differences.
The vorticity of the field $\boldsymbol{V} = (V_{x'}, V_{y'})$ was computed on a discretized square domain of $20 \times 20$ equally sized cells, each of side length $\Delta x' = \Delta y'= L / 20 = 5~\eta$ dimensionless units. The grid nodes are labelled by the pair by $(p,q)$, where $p=1,\dots,20$ increases along the horizontal ($x'$) axis and $q=1,\dots,20$ along the vertical ($y'$) axis. At each grid node, the scalar curl component normal to the plane is approximated using central finite differences as:
\[
\begin{split}
(\nabla \times \boldsymbol{V})_{p,q} \simeq \; & 
\frac{ V_{y'}(p+1,q) - V_{y'}(p-1,q) }{2\,\Delta x'} \\
& - \frac{ V_{x'}(p,q+1) - V_{x'}(p,q-1) }{2\,\Delta y'} \,.
\end{split}
\]

\subsubsection*{Computation of $\Phi_S$ on a discretized grid}
\label{sec:flux}
The flux defined in Eq.~\ref{eq:flux_surface} was evaluated as a line integral along a closed curve $l$, discretized into uniformly spaced segments:
\[
\Phi_S(R) \;\approx\; \sum_{m=1}^{N-1}
\Big[\boldsymbol{V}(x'_m, y'_m)\cdot \hat{\mathbf{n}}_m\Big]\;\Delta l_m ,
\]
where $\hat{\mathbf{n}}_m$ denotes the outward-pointing unit normal vector and $\Delta l_m$ the arc length of segment $m$. Two families of closed contours were considered: circular and square, both centered at $(x'_o, y'_o)$ and with increasing radius $R$. Field values at the contour were obtained by bilinear interpolation.
\paragraph{Circular contours.}
For a circle of radius $R$, the curve was parameterized as
\[
\begin{split}
x'_c(\theta) &= x'_o + R\cos\theta,\\
y'_c(\theta) &= y'_o + R\sin\theta ,
\end{split}
\quad \theta\in[0,2\pi],
\]
sampled at $N=100$ uniformly spaced points. The outward normal was computed from the radial direction:
\[
\hat{\mathbf{n}}_m(\theta) =
\frac{\big(x'_c(\theta)-x'_o,\; y'_c(\theta)-y'_o\big)}
{\sqrt{\big(x'_c(\theta)-x'_o\big)^2+\big(y'_c(\theta)-y'_o\big)^2}} .
\]
The arc length between consecutive points was approximated as
\[
\Delta l_m = \sqrt{\big[x'_c(\theta_{m+1})-x'_c(\theta_m)\big]^2
+\big[y'_c(\theta_{m+1})-y'_c(\theta_m)\big]^2} .
\]
The flux was then computed as
\[
\Phi_S(R)\approx\sum_{m=1}^{N-1}
\Big[\boldsymbol{V}\big(x'_c(\theta_m),y'_c(\theta_m)\big)
\cdot\hat{\mathbf{n}}(\theta_m)\Big]\,\Delta l_m .
\]
\paragraph{Square contours.}
Each side of the square (half-side $R$) was sampled with $N=25$ equidistant points. The point coordinates were
\[
\begin{split}
\text{Left:}\quad & x'=x'_o - R,\;\; y'\in[y'_o - R,\,y'_o + R],\\
\text{Top:}\quad & y'=y'_o + R,\;\; x'\in[x'_o - R,\,x'_o + R],\\
\text{Right:}\quad & x'=x'_o + R,\;\; y'\in[y'_o + R,\,y'_o - R],\\
\text{Bottom:}\quad & y'=y'_o - R,\;\; x'\in[x'_o + R,\,x'_o - R],
\end{split}
\]
with bilinear interpolation at each $(x'_m, y'_m)$. The unit normals, constant along each side, were
\[
\hat{\mathbf{n}}=
\begin{cases}
(\pm 1,0), & \text{Top, Bottom},\\
(0,\pm 1), & \text{Right, Left},
\end{cases}
\]
and the arc length per segment was $\Delta l = \frac{2R}{N}$.
Hence, the total flux was obtained as
\[
\Phi_S(R)\approx
\sum_{\text{sides}}\sum_{m=1}^N
\Big[\boldsymbol{V}(x'_m,y'_m)\cdot\hat{\mathbf{n}}\Big]\,\Delta l .
\]

\subsubsection*{Computation of $\Phi_V$ on a discretized grid}
\label{sec:divergence}
The volume flux in Eq.~\ref{eq:flux_volume} was computed as the surface integral of the divergence, evaluated on the discretized grid using finite differences. For each cell node $(p,q)$ in the region $S$ enclosed by the contour of radius $R$, the divergence was approximated as
\[
\begin{split}
\nabla\!\cdot\!\boldsymbol{V}(p,q) \approx
\frac{V_{x'}(p,q{+}1)-V_{x'}(p,q{-}1)}{2\,\Delta x'}+\\
\frac{V_{y'}(p{+}1,q)-V_{y'}(p{-}1,q)}{2\,\Delta y'},    
\end{split}
\]
with central differences applied to interior cells. On the boundary, forward or backward differences replaced the central method, e.g.,
\[
\begin{split}
\nabla\!\cdot\!\boldsymbol{V}(p,q) \approx
\frac{V_{x'}(p,q)-V_{x'}(p,q{-}1)}{\Delta x'}+\\
\frac{V_{y'}(p,q)-V_{y'}(p{-}1,q)}{\Delta y'}.
\end{split}
\]
The total flux was then obtained as
\[
\Phi_V(R)\approx
\sum_{(p,q)\in S}\nabla\!\cdot\!\boldsymbol{V}(p,q)\,\Delta S,
\]
where $\Delta S=\Delta x'\,\Delta y'$ is the area of a grid cell. Finally, after computing $\Phi_S$ and $\Phi_V$ for multiple $R$ values, their linear relationship was quantified using the Pearson correlation coefficient~\cite{pearson1896}.

\subsubsection*{Computation of the scalar potential \(\phi\)}\label{sec:potential}

The potential function 
\(\phi\) is obtained by numerically solving the relation \(\nabla^2 \phi = -\nabla \cdot \mathbf{V}\) under the assumption of an irrotational field,
\(\nabla \times \mathbf{V} = 0\).
To compute the empirical potential, Dirichlet boundary conditions are imposed by
setting \(\phi = 0\) on all boundary nodes of the grid. The divergence of \(\mathbf{V}\) is approximated using central finite differences, and the resulting equation is solved using the Jacobi iterative method \cite{saad2003}. For a grid cell with node indices \((p,q)\) and grid spacing \(h\), the discretized equation is:
\begin{equation*}
\begin{split}
\phi^{(k+1)}(p,q) = \frac{1}{4} \Big[ & \phi^{(k)}(p-1,q) + \phi^{(k)}(p+1,q) \\
& + \phi^{(k)}(p,q-1) + \phi^{(k)}(p,q+1) \\
& + h^2 (\nabla \cdot \mathbf{V})_{p,q} \Big]
\end{split}
\end{equation*}
The computation proceeds iteratively by updating the potential values across the grid until convergence is reached. The process stops when the maximum difference between consecutive iterations falls below a tolerance of \(10^{-6}\).

\section*{Data availability}
An illustrative toy dataset, generated to allow readers to execute and inspect the analysis code end to end, is available at (\url{https://github.com/paularshz/field_theory_in_football}); this dataset is provided solely for code demonstration and does not reproduce the statistical properties of the real tracking data used in this study. All raw data analyzed in the manuscript and its Supplementary Information were obtained under license from the proprietary data source LaLiga/Mediacoach and cannot be redistributed by the authors. Researchers wishing to apply the methodology described here to openly available data may consider high-resolution tracking datasets distributed by providers such as SkillCorner or Metrica Sports.
\section*{Code availability}
Custom MATLAB code is available on GitHub (\url{https://github.com/paularshz/field_theory_in_football}). 
Data and code (v1.0.0). Zenodo. \url{https://doi.org/10.5281/zenodo.18594252} (2026).

\section{References}
\bibliographystyle{naturemag}
\bibliography{ref}

\begin{acknowledgments}
J.M.B., J.H.M. and P.R.-S. acknowledge the support by the Agencia Estatal de Investigaci\'on of the Ministerio de Ciencia, Innovaci\'on y Universidades (MICIU/AEI/10.13039/501100011033), Spain under grant PID2023-147827NB-I00.
J.J.R.  was partially supported by the Spanish State Research Agency
(MICIU/AEI/10.13039/501100011033) and FEDER (UE) under projects COSASTI (PID2024-157493NB-C22), and the Mar{\'\i}a de Maeztu project CEX2021-001164-M.
\end{acknowledgments}
\section*{Author Contributions}
P.R.-S. J.J.R, J.M.B., and J.H.M. equally collaborated in the conception and supervision of the study, simulation experiments, figure preparation, result analysis, and the writing and revision of the manuscript. R.L-C, R.R were responsible of data extraction and data curation, and revised the manuscript with the rest of authors.
\section*{Competing interests}
The authors declare no competing interests.
\section*{Additional information}
Supplementary Information is available for this paper.

\end{document}